\documentclass[12pt]{article}
\usepackage{a4wide,theorem,amsfonts,amstext}
{\theoremstyle{break}
\theorembodyfont{\rmfamily}
\newtheorem{theorem}{Theorem}[section]

\newtheorem{lemma}{Lemma}[section]

}
\makeatletter
\@addtoreset{equation}{section}
\makeatother
\renewcommand{\thefootnote}{\fnsymbol{footnote}}

\newcommand{\be}{\begin{equation}}
\newcommand{\ee}{\end{equation}}
\newcommand{\ie}{\emph{i.e.}}

\newcommand\g{\ensuremath{\mathcal{G}}}

\begin{document}

\thispagestyle{empty}

\begin{flushright}

DAMTP/97--113~,~FTUV/97--46~,~IFIC/97--76

\end{flushright}

\vspace*{0.4cm}

\begin{center}

\begin{Large}

{\bf Effective actions, relative cohomology \\
and Chern Simons forms}

\end{Large}

\vspace*{0.4cm}

\begin{large}

J. A. de Azc\'arraga$^1$\footnote{E.mails: azcarrag@lie1.ific.uv.es,
a.j.macfarlane@damtp.cam.ac.uk, pbueno@lie.ific.uv.es},
A. J. Macfarlane$^{2*}$
and
J. C. P\'erez Bueno$^{1*}$
\end{large}

\vspace*{0.4cm}

$^1$ {\it Departamento de F\'{\i}sica Te\'orica and IFIC,
E-46100 Burjassot (Valencia), Spain}
\\[0.3cm]
$^2$ {\it Department of Applied Mathematics and Theoretical Physics,\\
Silver St., Cambridge, CB3 9EW, UK}

\end{center}

\setcounter{footnote}{0}
\renewcommand{\thefootnote}{\arabic{footnote}}

\begin{abstract}
The explicit expression of all the WZW effective actions for a simple group
$G$ broken down to a subgroup $H$ is established in a simple and
direct way, and the formal similarity of these actions to the
Chern--Simons forms is explained. Applications are also discussed.
\end{abstract}

\section{Introduction}
\label{eff.sec.1}
Recently it has been shown by D'Hoker and Weinberg \cite{DHo.Wei:94,DHo:95}
that the most general effective action of Wess--Zumino--Witten (WZW) type,
with a compact symmetry group $G$ broken down to a subgroup $H$, is given by
the non-trivial de Rham cocycles on the homogeneous coset manifold $G/H$, and a
cohomological descent-like procedure has been used in \cite{DHo:95} to obtain
explicit expressions for the lower order examples.
Motivated by this work \cite{DHo.Wei:94,DHo:95} and our own on the properties
of symmetric invariant tensors on simple algebras \cite{Azc.Mac.Mou.Bue:97}, we
look here at the problem of finding all the invariant effective actions of
WZW type in terms of the cohomology of the Lie algebra
\g\ \emph{relative} to a subalgebra ${\cal H}$ \cite{Che.Eil:48}.
By exploiting this (equivalent) point of view, we are able to find a general
formula for WZW type actions on $G/H$ for any
compact, connected and simply connected simple Lie group $G$ 
(the case of semisimple $G$ may be reduced to it) 
and for arbitrary spacetime dimensions.

The structure of phenomenological Lagrangians and nonlinear realizations was
elucidated thirty years ago \cite{Wei:68}.
Their relation to the standard Wigner little
group construction, which is the result of parametrising
the coset $K\equiv G/H=\{gH\vert g\in G\}$
in terms of the Goldstone coordinates
$\varphi^a\ (a=1,\dots,\text{dim}\,K)$, was emphasised in \cite{Sal.Str:69}.
Indeed, for the left action of a global transformation $g\in G=KH$ on 
the coset space $K$,
$g:\varphi^a\mapsto \varphi'^a$, we find
\be
g u(\varphi) = u(\varphi') h(g,\varphi)\quad,
\label{eff1.1}
\ee
where $h(g,\varphi)\in H$ is simply an element of the `little group' of
the coset reference point.
Geometrically, $u(\varphi)$
({\it e.g.}, $\exp{\varphi^a T_a}$) may be viewed as
a section of the bundle $G(H,K)$; since $gu(\varphi)$ and $u(\varphi')$
belong to the same fibre, eq. (\ref{eff1.1}) follows\footnote{Notice that 
the left action of $G$ on $G$ and on $K$ (defined by the left cosets
$gH$) and the right action of $H$ on $G$ are both compatible
with the bundle projection.}.

The key idea of the standard nonlinear realizations in general 
is that we may write
\be
u^{-1} d u = (u^{-1} d u)_{{\cal H}} + (u^{-1} d u)_{{\cal K}} \equiv
\Omega_{{\cal H}}^\alpha T_\alpha + \Omega_{{\cal K}}^a T_a
\quad.
\label{eff1.2}
\ee
As is well known, \cite[I, p.103]{Kob.Nom:63} the 
${\cal H}$-valued component of the canonical left invariant (LI) form 
$\omega$ on $G$ determines a
LI connection on $G$; the first term in (\ref{eff1.2}) is
its pull-back to $K$ by the section $u(\varphi)$, and 
hence transforms as a
connection. In contrast, $\Omega_{{\cal K}}$ transforms 
tensorially through $h(g,\varphi)\in H$ under a
left transformation of $G$, its elements operating linearly for $H\subset G$
and nonlinearly for those in $G\backslash H$.
Hence any $H$--invariant expression made out of $\Omega_{{\cal K}}$
will also be invariant under the whole $G$ and is a candidate for an invariant
nonlinear Lagrangian. However, as emphasised in \cite{DHo.Wei:94},
there are also invariant actions which do not come from an invariant
Lagrangian density.
These are generically referred to as WZW actions \cite{Wes.Zum:71},
also discussed in the context of nonlinear sigma models
in \cite{wzgen}.
In Witten's derivation \cite{Wes.Zum:71} of the simplest WZW term 
for $G=SU(2)$ ($H$=$e$) and $D$=2 spacetime $M\sim S^2$, the 
fields $g(x)$ are extended by means
of an interpolating field $g(x,\lambda)$ ($g(x,0)=0,\ g(x,1)=g(x)$) and the
WZW action is given by
\be
I_{WZW} =\int_B d^2 x d\lambda \epsilon^{\mu \nu} \text{Tr} (g^{-1} 
\;{\partial g/ \partial\lambda}\, W_\mu W_\nu )
\label{eff1.3}
\ee
where $W_\mu = g^{-1} \partial_\mu g$ and $\partial B=M$;
the construction uses $\pi_2(G)=0$ and
$\pi_3(G)=\mathbb{Z}$ (which hold for any simple Lie group).
Similar considerations can be made for higher $D$, where
the existence of a WZW term requires in particular \cite{Azc.Izq.Mac:90}
that there is a non-trivial $(D+1)$-cocycle (form on $G$) 
for the Chevalley-Eilenberg \cite{Che.Eil:48} (CE) cohomology.

When $H\ne e$, the  mappings $\varphi^a:M\to G/H$ are the Goldstone fields,
suitable extended to the analogous of $B$ above \cite{DHo.Wei:94}.
The construction of D'Hoker and Weinberg shows that the WZW actions (\ie,
invariant actions associated with non-invariant spacetime Lagrangian densities)
on the coset $K$ are given by non-trivial De Rham cocycles on $K$,
\ie, by closed non-exact forms on $G/H$ 
. The result of \cite{DHo.Wei:94,DHo:95} may be reformulated by
stating that the WZW actions are classified by the non-trivial cocycles of the
relative algebra cohomology $H_0(\g,{\cal H};\mathbb{R})$
(for $H=e$, $H_0(\g,e;\mathbb{R})=H_0(\g;\mathbb{R}))$, and
this approach will lead us to a general expression for them.
We shall restrict ourselves here to the ungauged case, and
will not discuss the (related) problem of gauging the WZW actions 
\cite{wzgen}.

This paper is organised as follows: in Sec. \ref{eff.sec.2} we review briefly
the forms on coset manifolds and the relative Lie algebra cohomology.
Sec. \ref{eff.sec.3} is devoted to finding an explicit general formula for the
non-trivial cocycles on $G/H$ for a simple group, and in Sec. \ref{eff.sec.4}
we illustrate our result with applications.
The formal connection between the cocycles on $G/H$ and the expression
for the Chern-Simons forms is exhibited in Sec. \ref{eff.sec.5}, where the
relation between the two is clarified.
The indices are as follows: $i,j,...$ refer
to $G$ (or its algebra \g), $i=1,\dots,$dim$\,G$; $\alpha,\beta,...$ to
the subgroup $H$ (${\cal H}$), $\alpha=1,...,$dim$\,H$, and the 
indices $a,b,...$ parametrise the coset $K$ (or the vector space 
${\cal K}={\cal G}/ {\cal H}$), $a=1,...,$dim$\,K$.

\section{Forms on cosets and relative algebra cohomology}
\label{eff.sec.2}

{}From the point of view of physical applications, Lie algebra \g\ cohomology
groups are most conveniently described \cite{Che.Eil:48} in terms of forms on
the associated simply connected group manifold $G$.
For the trivial representation $\rho(\g)=0$, the cohomology groups
$H_0^q(\g,\mathbb{R})$ are characterised by a) closed and b) (say) left
invariant (LI) $q$-forms on $G$ (the $q$-cocycles) modulo those which are 
the exterior derivative $d$ of a LI form (the coboundaries).
Let ${\omega^i}$ be a basis of LI one-forms on $G$ so that
\be
d\omega^i = -{1\over 2} C^i_{jk} \omega^j \wedge \omega^k
\quad (i,j,k=1,...,{\mbox{dim}}\,\g)\quad,\quad
\text{or}
\quad
d\omega = - \omega \wedge \omega \quad
\label{eff2.0}
\ee
($\omega$ is the \g-valued canonical form on $G$).
Then the LI $q$-forms $\Omega$ on $G$ may be written as
\be
\Omega= {1\over q!} \Omega_{i_1\dots i_q}
                        \omega^{i_1} \wedge \dots \wedge \omega^{i_q}\quad,
\label{eff2.1}
\ee
Then, those which determine Lie algebra $q$-cocycles satisfy the condition
\be
(s\Omega)_{i_1\dots i_{q+1}} = - {1\over 2} {1 \over (q-1)!} C^{l}_{[i_1 i_2}
        \Omega_{l i_3 \dots i_q]} =0\quad,
\label{eff2.2}
\ee
where, in the CE formulation, the coboundary
operator $s$ may be identified with the exterior derivative $d$.

In the language of forms the {\it relative cohomology} with
respect to a subalgebra ${\cal H}\subset\g$ is associated with the notion of
projectability of forms on $G$ to the coset manifold $K=G/H$.
This notion, which plays an essential r\^ole in the Chern-Weil theory of
characteristic classes, actually means that there is a unique form
$\bar\Omega$ on $G/H$ such that $\pi^*(\bar\Omega)=\Omega$, where $\pi^*$ is
the pull-back of the canonical projection $\pi:G\to G/H$.
A $q$-form $\Omega$ is projectable if (see, \emph{e.g.}
\cite[II, p. 294]{Kob.Nom:63})
\be
\Omega(X_1,\dots,X_q)=0\ \text{if any}\ X\in {\cal H}  \quad,
\label{eff2.6}
\ee
\be
(L_{X_\alpha} \Omega)(X_1,\dots,X_q) =-\sum_{s=1}^q
\Omega(X_1,\dots,[X_{\alpha},X_s],\dots,X_q) =0 \quad,
\label{eff2.7}
\ee
\ie, if $\Omega$ is `orthogonal' to ${\cal H}$ [(\ref{eff2.6})] and it is
invariant under the right action of $H$ [(\ref{eff2.7})]. In (\ref{eff2.7}),
$L_X$ is the Lie derivative with respect to the LI vector
field $X$; on LI forms, $L_{X_i}\omega^j=-C^i_{jk}\omega^k$.
As a result, a $q$-form $\Omega$ is a non-trivial $q$-cocycle for the
relative Lie algebra cohomology $H_0^q(\g,{\cal H};\mathbb{R})$
\cite{Che.Eil:48} if {\it a}) it is LI and closed, eq. (\ref{eff2.2}) 
(\ie, it is a $q$-cocycle in $Z_0^q(\g;\mathbb{R})$); {\it b})
it is projectable and {\it c}) it is not the 
exterior derivative of a LI, projectable form (in which case
it would be a coboundary). Our task is now to find the closed
forms on the coset manifold $G/H$, parametrised by 
$\varphi^a$, which are non-trivial cocycles in $H_0(\g,{\cal H};\mathbb{R})$.
These will define effective actions of WZW type once they are pulled 
back to an enlarged spacetime manifold of the appropriate dimension.

\section{An explicit formula for the cocycles on $G/H$}
\label{eff.sec.3}

As is well known, the non-trivial primitive cocycles
(\ie, that are not the product of other cocycles) 
on a simple group $G$ are all of odd order
\cite{Car:36}
$(2m_s-1)$, $s=1,\dots,l$ where $l$ is the rank of \g. They are 
associated with the $l$ primitive invariant symmetric tensors 
$k_{i_1...i_{m_s}}$ of order $m_s$
(and Casimir operators of the same order) which may be constructed on \g\
\cite{Rac:50},
the properties of which have been studied recently \cite{Azc.Mac.Mou.Bue:97}.
Given such a tensor $k_{i_1\dots i_m}$, the \g-(2$m$--1)-cocycle
$\Omega^{(2m-1)}$, or simply $\Omega$,
is a form on $G$ $\Omega \propto
k_{l_1\dots l_{m-1} s} d\omega^{l_1}\wedge\dots\wedge d\omega^{l_{m-1}}
\wedge \omega^{s}$, so that its coordinates are proportional to
\be
\Omega_{i_1 \dots i_{2m-2} s} \propto
k_{l_1\dots l_{m-1} [s} C^{l_1}_{i_1 i_2} \dots C^{l_{m-1}}_{i_{2m-3}
i_{2m-2}]}
\label{eff3.1}
\ee
(any non-primitive terms in $k$ do not contribute to (\ref{eff3.1});
see Cor.3.1 in \cite{Azc.Mac.Mou.Bue:97}).
Due to the full antisymmetry of the structure constants, any semisimple group
is reductive,
$C_{\alpha\beta}^a=0$, $C_{\alpha a}^{\beta}=0$ ($\alpha,\ \beta$ in
${\cal H}$, $a,\ b$ in ${\cal K}$),
\ie,
$[{\cal H},{\cal H}]\subset {\cal H},\
[{\cal H},{\cal K}]\subset {\cal K},\
[{\cal K},{\cal K}]\subset \g$.
The cocycles on $G/H$ are the projectable cocycles on $G$.
To find general expressions for them for any simple $G$ consider the
(2$m$-1)-form $\Omega_{(p)}$ (cf. (\ref{eff3.1})),
\be
\begin{array}{l}
\Omega_{(p)} = k_{\alpha_1\dots \alpha_{p-1} i_{p}\dots i_{m-1} b}
C^{\alpha_1}_{a_1 a_2} \dots C^{\alpha_{p-1}}_{a_{2p-3} a_{2p-2}}
C^{i_{p}}_{a_{2p-1} a_{2p}} \dots C^{i_{m-1}}_{a_{2m-3} a_{2m-2}}
\\[0.3cm]
\hfill \cdot \omega^{a_{1}}\wedge\dots\wedge
\omega^{a_{2p}}\wedge\omega^{a_{2p+1}}
\wedge\dots\wedge
\omega^{a_{2m-2}}\wedge \omega^{b}
\quad.
\end{array}
\label{eff3.2}
\ee
This form clearly satisfies the condition (\ref{eff2.6})
($i_{X_\alpha}\Omega_{(p)} =0$) since $\omega^{a}(X_\alpha)=0\
\forall\ a$ in ${\cal K}$ and $\alpha$ in ${\cal H}$.
The proof that $L_{X_\alpha} \Omega_{(p)}=0$, eq. (\ref{eff2.7}) follows from
$L_{X_\alpha} \omega^a = -C_{\alpha b}^a \omega^b$ and the fact that the
constants preceding
$\omega^{a_1} \wedge \dots \wedge \omega^{a_{2m-2}} \wedge \omega^b$ in
(\ref{eff3.2}) may be viewed as products of the invariant polynomials $k,\ C$
to which we may apply the following

\begin{lemma}
\label{eff.lem.3.1}
Let $k_{i_1\dots i_{n+1}}$ and $k'_{j_1\dots j_{m+1}}$ be
two invariant tensors on $G$ (symmetry is not required here).
Then,
\be
L_\alpha \left( k_{a_1 \dots a_n \beta} k'_{\beta b_1 \dots b_m}
\omega^{a_1}\otimes \dots \otimes \omega^{a_n} \otimes
\omega^{b_1}\otimes \dots \otimes \omega^{b_m} \right) = 0
\label{eff.lem.3.1a}
\ee
and
\be
L_\alpha \left( k_{a_1 \dots a_n j} k'_{j b_1 \dots b_m}
\omega^{a_1}\otimes \dots \otimes \omega^{a_n} \otimes
\omega^{b_1}\otimes \dots \otimes \omega^{b_m} \right) = 0
\label{eff.lem.3.1b}
\ee
{\it Proof:}
By using the ($G$)-invariance of $k$ we obtain
$$
C^i_{\alpha a_1} k_{i a_2\dots a_n \beta}
+\dots +
C^i_{\alpha a_n} k_{a_1 \dots a_{n-1} i \beta}
+
C^i_{\alpha \beta} k_{a_1 \dots a_n i} =0 \quad.
$$
Now, using the fact that the coset is reductive, we get
$$
C^a_{\alpha a_1} k_{a a_2\dots a_n \beta}
+\dots +
C^a_{\alpha a_n} k_{a_1 \dots a_{n-1} a \beta}
+
C^\gamma_{\alpha \beta} k_{a_1 \dots a_n \gamma} =0
$$
and similarly for $k'$. Thus, 
$$
\begin{array}{l}
\left(C^a_{\alpha a_1} k_{a a_2\dots a_n \beta}
+\dots +
C^a_{\alpha a_n} k_{a_1 \dots a_{n-1} a \beta} \right)
k'_{\beta b_1 \dots b_m}
\\
\displaystyle
+
k_{a_1 \dots a_n \beta} \left(
C^a_{\alpha b_1} k'_{\beta a b_2\dots b_m}
+\dots +
C^a_{\alpha b_m} k'_{\beta b_1 \dots b_{m-1} a} \right)
\\
=
-C^\gamma_{\alpha \beta} k_{a_1 \dots a_n \gamma} k'_{\beta b_1 \dots b_m}
- k_{a_1 \dots a_n \beta} C^\gamma_{\alpha \beta} k'_{\gamma b_1 \dots b_{m}}
=0
\end{array}
$$
from which (\ref{eff.lem.3.1a}) follows.
Eq. (\ref{eff.lem.3.1b}) is deduced from the fact that
$k_{i_1 \dots i_n j} k'_{j j_1 \dots j_{m}}$ is an invariant tensor on $G$ so
that
$$
\sum_{s=1}^n C^k_{\alpha i_s} k_{i_1 \dots \hat{i_s} k \dots i_n j}
k'_{j j_1 \dots j_{m}}
+
\sum_{t=1}^m C^k_{\alpha j_t} k_{i_1 \dots i_n j}
 k'_{j j_1 \dots \hat{j_t} k \dots j_{m}} =0\quad.
$$
Setting $i_s=a_s$ and $j_t=b_t$ and using again the reductive property 
we obtain (\ref{eff.lem.3.1b}), {\it q.e.d.}.
\end{lemma}

Since the cocycles on the coset manifold are LI closed forms on $K$, we look
now for a closed form.
Since $L_\alpha\Omega_{(p)} = (i_{X_\alpha} d + d i_{X_\alpha}) \Omega_{(p)} =
i_{X_\alpha} d\Omega_{(p)} =0$, when computing $d\Omega_{(p)}$ we may ignore
the $\omega^\alpha$ components.
A straightforward if somewhat lengthy calculation, which uses the
Jacobi identity and the fact that the coset is reductive shows that
\be
d\Omega_{(p)} = -{1\over 2} \Pi_{(p-1)} +{2m-p\over 2p} \Pi_{(p)} \quad,
\label{eff3.4}
\ee
where the $2m$-forms $\Pi_{(p-1)}$ and $\Pi_{(p)}$ are given by
\be
\Pi_{(p)} = k_{\alpha_1\dots \alpha_p i_{p+1}\dots i_m}
C^{\alpha_1}_{a_1 a_2} \dots C^{\alpha_p}_{a_{2p-1} a_{2p}}
C^{i_{p+1}}_{a_{2p+1} a_{2p+2}} \dots C^{i_{m}}_{a_{2m-1} a_{2m}}
\omega^{a_1}\wedge\dots\wedge \omega^{a_{2m}}
\quad.
\label{eff3.5}
\ee
For $p=0$ we find that that $\Pi_{(0)}=0$ due to the \g-invariance of
$k_{i_1\dots i_m}$.
Let us now define a new (2$m-1$)-form $\bar\Omega$ by
\be
\begin{array}{c}
\displaystyle
\bar\Omega=\sum_{s=1}^m \alpha_m(s) \Omega_{(s)}\quad,
\\[0.3cm]
\displaystyle
\alpha_m(s)\equiv \prod_{r=1}^{s-1} {2m-r\over r} \quad,\quad
\alpha_m(1)\equiv 1\quad,\quad
\alpha_m(m)={(2m-1)!\over (m-1)! m!}\quad.
\end{array}
\label{eff3.6}
\ee
Then we find from (\ref{eff3.4})
\be
\begin{array}{rl}
d\bar\Omega &
=
\displaystyle
\sum_{s=1}^m \alpha_m(s) \left ( - {1\over 2}\Pi_{(s-1)} + {2m-s\over 2s}
\Pi_{(s)} \right)
\\[0.4cm]
&
=
\displaystyle
-{1\over 2} \left (\sum_{s=0}^{m-1} \alpha_m(s+1) \Pi_{(s)} -
\sum_{s=1}^m \alpha_m(s)
{2m-s\over s} \Pi_{(s)}
\right)
\\[0.4cm]
&
=
\displaystyle
-{1\over 2} \left (\sum_{s=1}^{m-1} \alpha_m(s+1) - \alpha_m(s) {2m-s\over s}
\right)  \Pi_{(s)}
+{1\over 2} \alpha_m(m) \Pi_{(m)}
=
\displaystyle
{1\over 2} \alpha_m(m) \Pi_{(m)}
\end{array}
\label{eff3.7}
\ee
Hence $d\bar\Omega$ will be zero if $\Pi_{(m)}=0$ \ie, if
$k_{\alpha_1\dots \alpha_m}=0$.
Thus, we have proven the following

\begin{theorem}
\label{eff.th.3.1}
Let $G$ be a simple, simply connected group and $H$ a closed subgroup.
A primitive non-trivial (2$m-1$)--cocycle in
$H^{(2m-1)}(\g,{\cal H};{\mathbb R})$ is represented by the closed
(2$m$--1)-form $\bar\Omega$ on the coset $G/H$ given in (\ref{eff3.6}).
This form is defined through (\ref{eff3.5})
by a polynomial of order $m$ on \g\ which vanishes on ${\cal H}$,
\ie, when \emph{all} its indices take values in ${\cal H}$.
\end{theorem}

\section{Applications}
\label{eff.sec.4}

As an application of our general formula (\ref{eff3.6}) let us find the
expression for the three- $(m=2)$ and five- $(m=3)$ cocycles.
Eq. (\ref{eff3.6}) gives
\be
\bar\Omega^{(3)} = \Omega_{(1)} + \alpha_2(2) \Omega_{(2)}
=(k_{i_1 a_3} C^{i_1}_{a_1 a_2} + 3 k_{\alpha_1 a_3} C^{\alpha_1}_{a_1 a_2})
\omega^{a_1} \wedge \omega^{a_2} \wedge\omega^{a_3}
\label{eff3.8}
\ee
\be
\begin{array}{c}
\bar\Omega^{(5)} = \Omega_{(1)} + \alpha_3(2) \Omega_{(2)} +
\alpha_3{(3)}\Omega_{(3)} =
\\[0.3cm]
(k_{i_1 i_2 a_5} C^{i_1}_{a_1 a_2} C^{i_2}_{a_3 a_4}
+ 5 k_{\alpha_1 i_2 a_5} C^{\alpha_1}_{a_1 a_2} C^{i_2}_{a_3 a_4}
+ 10 k_{\alpha_1 \alpha_2 a_5} C^{\alpha_1}_{a_1 a_2} C^{\alpha_2}_{a_3 a_4}
) \omega^{a_1} \wedge \dots \wedge\omega^{a_5}
\quad.
\end{array}
\label{eff3.9}
\ee
These two expressions have also been derived by D'Hoker by a lengthier
cohomological descent-like procedure \cite{DHo:95}, involving the
consideration of non-trivial representations $\rho$ of $\g$.
To exhibit the computational convenience
of the general formula (\ref{eff3.6}), we give one further example,
the seven-cocycle,
\be
\begin{array}{rl}
\bar\Omega^{(7)} & =
\Omega_{(1)} + \alpha_4(2) \Omega_{(2)} +
\alpha_4{(3)}\Omega_{(3)} + \alpha_4{(4)}\Omega_{(4)} =
\\[0.3cm]
& = (k_{i_1 i_2 i_3 a_7} C^{i_1}_{a_1 a_2} C^{i_2}_{a_3 a_4} C^{i_3}_{a_5 a_6}
+ 7 k_{\alpha_1 i_2 i_3 a_7} C^{\alpha_1}_{a_1 a_2} C^{i_2}_{a_3 a_4}
C^{i_3}_{a_5 a_6}
\\[0.3cm]
& +
21 k_{\alpha_1 \alpha_2 i_3 a_7} C^{\alpha_1}_{a_1 a_2} C^{\alpha_2}_{a_3 a_4}
C^{i_3}_{a_5 a_6}
+ 35 k_{\alpha_1 \alpha_2 \alpha_3 a_7} C^{\alpha_1}_{a_1 a_2}
C^{\alpha_2}_{a_3 a_4} C^{\alpha_3}_{a_5 a_6}
)
\omega^{a_1} \wedge \dots \wedge\omega^{a_7} \quad.
\end{array}
\label{eff3.10}
\ee
Of course, the existence of these cocycles depends on the existence of
invariant polynomials of the appropriate degree which are
zero on ${\cal H}$.
These are all known for all simple algebras (see
also \cite{Azc.Mac.Mou.Bue:97} in this respect); in particular all three
exist for $\g=su(n)$, $n\ge m$.

Let us consider now some specific examples.

{\it a}) ($SU(n)/SU(m)$ {\it cosets})

Consider the case $K=SU(3)/SU(2)\sim S^5$.
To construct a 5-cocycle on $S^5$ we need a $3rd$-order polynomial vanishing
on $SU(2)$.
This is provided by the $d_{i_1 i_2 i_3}$ polynomial which satisfies
$d_{\alpha_1 \alpha_2 \alpha_3}=0\ \forall \alpha \in SU(2)$.
Similarly for the case $K=SU(4)/SU(2)$ we have two invariant polynomials of
($3rd$ and $4th$ order)  vanishing on $SU(2)$, which give rise to the 5- and
7-cocycles respectively, etc. For the case $m>2$, we can always
construct symmetric invariant polynomials on $su(n)$ which are zero on
$su(m)$ (see \cite{Azc.Mac.Mou.Bue:97}).

{\it b}) ({\it Symmetric cosets})

The simplest cases in which formula (\ref{eff3.6}) gives rise to a
non-trivial result are furnished by \emph{symmetric cosets}
$G/H$ ($[{\cal K},{\cal K}]\subset {\cal H}$).
Such examples, when they exist, are simple because then all terms
in (\ref{eff3.6}) have the same structure since $C_{ab}^c$=0. 
They require $k_{\alpha_1\dots\alpha_m}$=0 and that the components 
$k_{\alpha_1\dots\alpha_{m-1}a}$ do not all vanish.
For instance, for $G=SU(n)$, $m=3$ and $d_{\alpha_1\alpha_2\alpha_3}$=0,
the five-cocycle becomes proportional to
\be
d_{\alpha_1 \alpha_2 a_1} C^{\alpha_1}_{a_2 a_3} C^{\alpha_2}_{a_4 a_5}
\omega^{a_1} \wedge \dots \wedge \omega^{a_5}
\quad.
\label{eff.s.ex1}
\ee
Symmetric spaces in which the $d_{\alpha_1 \alpha_2 \alpha_3}$ vanish and the
$d_{\alpha_1 \alpha_2 a_1}$ do not, are provided by the families
\cite[p. 518]{Hel:78}
$SU(n)/SO(n)$ and $SU(2n)/Sp(2n)$.

The simplest case, one which leads directly to a Wess-Zumino term in a four
dimensional field theory, is that of $SU(3)/SO(3)$.
To clarify this, we work with the generators $T_A ={1\over 2} \lambda_A$ of
$su(3)$, where the $\lambda_A$ are the set of standard Gell-Mann matrices.
For the $so(3)$ generators, we take $\alpha\in\{2,5,7\}$ and,
since $C_{2 5}^7= {1\over 2}$,
$[\lambda_\alpha,\lambda_\beta]=i\epsilon_{\alpha\beta\gamma}\lambda_\gamma$.
Then the coset indices $a\in\{1,3,4,6,8\}$.
To see that it is correct to write $SU(3)/SO(3)$, we note that reduction of
the octet $SU(3)$ with respect to $SO(3)$ produces $j=1$ and $j=2$
$SO(3)$-multiplets, and we can argue that integral $j$ values only arise in
the reduction of triality zero $SU(3)$ representations.
Explicitly, we can show that, for $c\in \mathbb{R}$ 
\be
{i\over\sqrt 2} c (\lambda_1 - i\lambda_3)\quad,\quad
{i\over\sqrt 2} c (\lambda_4 + i\lambda_6) \quad,\quad
c\lambda_8\quad,
\quad{i\over\sqrt 2} c (\lambda_4 - i\lambda_6) \quad,\quad
{-i\over\sqrt 2} c (\lambda_1 + i\lambda_3)\quad,
\label{eff.s.ex2}
\ee
are the standard Racah components $T_q$, $q=(2,1,0,-1,-2)$ of a rank $2$
tensor operator of $SO(3)$.
In this example, one can see by inspection of the $d$-tensor of $SU(3)$, that
the $d_{\alpha_1 \alpha_2 \alpha_3}$ all vanish, but there are eight non-zero
triples for which
$d_{\alpha_1 \alpha_2 a_1}\ne 0$.
It is in fact easy to see without explicit calculation that
(\ref{eff.s.ex1}) is a multiple of
$\omega^1\wedge\omega^3\wedge\omega^4\wedge\omega^6\wedge\omega^8$.

To discuss this and other examples involving $G=SU(4)$, it is most convenient 
to generalise the Gell-Mann $\lambda$--matrices from $SU(3)$ to $SU(4)$ in a 
fashion different from that in \cite{Azc.Mac.Mou.Bue:97, Hay:76}  
and to use the $d$ and $f$ tensors that follow from this new set.
Thus, set $\lambda_i = \pmatrix{ \sigma_i & 0 \cr 0 & 0}$, 
$\lambda_{i+12} = \pmatrix{ 0 & 0 \cr 0 & \sigma_i}$, 
$\lambda_8=\sqrt{1\over 2} \pmatrix {1 & 0 \cr 0 & -1}$, 
where $\sigma_i$ are the three Pauli matrices.
For $i=4$ to 7, we retain the $\lambda_i$ of $SU(3)$ so that for $i$=4 to 7
and 9 to 12, the $\lambda_i$ of \cite{Hay:76} are used.
In particular, $\lambda_3,\ \lambda_8,\ \lambda_{15}$ are diagonal, all 
$\lambda$'s are hermitian and $\lambda_i$ for $i\in$\{2,5,7,10,12,14\} are 
antisymmetric.
In fact we have only changed our choices of $\lambda_8$ and 
$\lambda_{15}$, so 
that very little further evaluation of $d$ and $f$ tensors is needed.

Consider first $G/H$=$SU(4)/SO(4)$ in which $SO(4)$ is generated by 
the set of six antisymmetric $\lambda$'s just mentioned, 
while the $SU(4)$ $d$-tensors 
are as tabulated in \cite{Azc.Mac.Mou.Bue:97}. The 
$d_{\alpha_1\alpha_2\alpha_3}$ do vanish, since 
they correspond to the trace 
of products of three antisymmetric matrices, while for many triples the 
$d_{\alpha_1\alpha_2 a_3}\ne 0$.
Thus a simple five-cocycle is allowed in this model with nine
 Goldstone fields.
We may contrast this with the {\it non}-symmetric reductive model 
$SU(4)/[SU(2)\times SU(2)]$ in which the the subgroup 
generators are $\lambda_i$ for $i\in$\{1,2,3,13,14,15\}.
The relevant $d_{\alpha_1\alpha_2\alpha_3}$ obviously vanish but, 
since the $C_{ab}^c$ are not all zero, all the three 
terms of (\ref{eff3.8}) survive giving a much more complicated 
Wess-Zumino term for this model with 15-6=9 Goldstone fields.

 Consider next the symmetric coset $SU(4)/S[U(2)\times U(2)]$, 
where ${\cal H}$ is larger than in the previous example, with the extra 
generator $\lambda_8$. The $d_{\alpha_1\alpha_2\alpha_3}$ do not all 
vanish now, and hence there are no five-cocycles of type 
(\ref{eff3.6}) \cite{DHo.Wei:94}.
Finally, consider the five-dimensional symmetric coset 
$SU(4)/Sp(4,{\mathbb R})$. A presentation
of $C_2=sp(4,{\mathbb R})$ in Cartan-Weyl form with positive 
roots $r_1=(1,-1)$, $r_2=(0,2)$, $r_3=(1,1)$, $r_4=(2,0)$ 
can be given in terms of the above $4\times 4$ 
$\lambda$-matrices of $SU(4)$.
Writing $\sqrt{2} E_{\pm\mu}=X_\mu \pm iY_\mu$ for $\mu=1, 2, 3, 4$ for the 
raising and lowering operators associated with the roots, the 
realisation is
\be
\begin{array}{c}
H_1=\lambda_3\quad,\quad X_4=\lambda_1\quad,\quad Y_4=\lambda_2\quad,
\\
H_2=\lambda_{15}\quad,\quad X_2=\lambda_{13}\quad,\quad Y_2=\lambda_{14}\quad,
\\
\sqrt{2} X_1=\lambda_{4}-\lambda_{11}\quad,\quad
\sqrt{2} Y_1=\lambda_{5}+\lambda_{12}\quad,
\\
\sqrt{2} X_3=\lambda_{6}+\lambda_{9}\quad,\quad
\sqrt{2} Y_3=-\lambda_{7}+\lambda_{10}\quad.
\end{array}
\ee
This enables an explicit check that the corresponding 
$d_{\alpha_1\alpha_2\alpha_3}$ are indeed all zero, an easy tabulation of 
$C_2$ structure constants $C_{\alpha \beta}^{\gamma}$,
$(1\le \alpha,\beta,\gamma\le 10)$, and the 
evaluation of the five-cocycle of the model.

There are no seven-dimensional symmetric cosets. Consider then
the case of the  9-cocycle determining a Wess-Zumino term in $D=8$
spacetime. For $G/H$ symmetric, all the coordinates of 
$\bar\Omega^{(9)}=\bar\Omega^{(9)}_{a_1 \dots a_9}
\omega^{a_1} \wedge \dots \wedge \omega^{a_9}$ become 
proportional to 
\be
k^{(5)}_{\alpha_1 \alpha_2 \alpha_3\alpha_4 [ a_1}
C^{\alpha_1}_{a_2 a_3} C^{\alpha_2}_{a_4 a_5} C^{\alpha_3}_{a_6 a_7}
C^{\alpha_4}_{a_8 a_9]}
\label{eff.s.ex5}
\ee
Let $G/H=SU(n)/SO(n)$ with $so(n)$ generated by 
the $\frac 12 n(n-1)\ n\ge 4$ imaginary antisymmetric $\lambda_\alpha$ 
matrices, the coset generators
$\lambda_a$ being real symmetric and traceless.
Then, if we take 
$k_{i_1\dots i_5}\sim \text{sTr}(\lambda_{i_1}\dots\lambda_{i_5})$ it is
obvious that $\text{sTr}(\lambda_{\alpha_1}\dots\lambda_{\alpha_5})$ is zero
but that 
$\text{sTr}(\lambda_{\alpha_1}\dots\lambda_{\alpha_4}\lambda_a)$ is not.
Hence we will get a Wess-Zumino term from 
\be
\bar\Omega^{(9)}_{a_1 \dots a_9}
\omega^{a_1} \wedge \dots \wedge \omega^{a_9}\quad,
\quad\bar\Omega^{(9)}_{a_1 \dots a_9}\propto
k^{(5)}_{\alpha_1 \alpha_2 \alpha_3 \alpha_4 [ a_1}
C^{\alpha_1}_{a_2 a_3} C^{\alpha_2}_{a_4 a_5} C^{\alpha_3}_{a_6 a_7}
C^{\alpha_4}_{a_8 a_9]} \quad .
\label{eff.s.ex8}
\ee
In (\ref{eff.s.ex8}) we may use equivalently $d^{(5)}$ 
for the $su(n)$ polynomial, (see \cite{Azc.Mac.Mou.Bue:97}).

\section{Relative cohomology and Chern-Simons forms}
\label{eff.sec.5}

Let us consider $\Omega_{(p)}$ in (\ref{eff3.2}) further.
First we introduce
\be
k_{\alpha_1 \dots \alpha_{p-1} i_p \dots i_{m-1} b}
={1\over m!} \text{sTr}
(\text{ad} X_{\alpha_1} \dots \text{ad} X_{\alpha_{p-1}}
\text{ad} X_{i_p} \dots \text{ad} X_{i_{m-1}} \text{ad} X_{b})
\label{eff4.1}
\ee
which arises by restricting the indices of the invariant symmetric
polynomial of order $m$ on \g, given by a symmetric trace,
to the appropriate values.
Next, we make the identifications 
\be
W^\alpha = - {1\over 2} C^{\alpha}_{a b} \omega^a \wedge \omega^b
\quad,\quad
{\cal U} = \omega\vert_{\cal K} \quad,\quad
({\cal U}\wedge{\cal U})^i \equiv
({\cal U}^2)^i = {1\over 2} C^{i}_{a b} \omega^a \wedge \omega^b \quad ,
\label{eff4.2a}
\ee
and ${\cal V}= \omega\vert_{\cal H}$,  where $\omega=u^{-1}du$. 
Thus, ${\cal V}$ 
determines the LI invariant ${\cal H}$-connection  
and $W$ the associated curvature,
$W=d{\cal V} + {\cal V}\wedge{\cal V}$ (which leads to $W^\alpha$ in
(\ref{eff4.2a}) using the Maurer-Cartan eqs.). Then, 
the form in (\ref{eff3.2}) may be rewritten as
\be
\Omega_{(p)} = (-1)^{p-1} {2^{m-1}\over m!}
\text{sTr} \{W^{p-1} ({\cal U}^2)^{m-p} {\cal U}\} \quad.
\label{eff4.2}
\ee
If we replace the $m!$ terms in sTr by the sum ${\cal S}$ over all possible
products (`words') which contain a total power $(p-1)$ $[(m-p)]$ of the
curvature $W$
[component ${\cal U}$], eq. (\ref{eff4.2}) may be written as
\be
\Omega_{(p)} =
{2^{m-1}\over (m-1)!}
(-1)^{p-1} (p-1)! (m-p)! \text{Tr}
\{ \mathcal{S} \left [ (W^{p-1} ({\cal U}^2)^{m-p}) \right ] {\cal U}\}
\quad.
\label{eff4.3}
\ee
As a result, the general expression
(\ref{eff3.6}) for the (2$m-1$)-cocycle on the coset $K$ may be rewritten as
\be
\bar\Omega^{(2m-1)} =
{2^{m-1}\over (m-1)!}
\sum_{p=0}^{m-1} (-1)^p
(p)! (m-p-1)! \alpha_m(p+1) \text{Tr}
\{ \mathcal{S} \left [ (W^{p} ({\cal U}^2)^{m-p-1}) \right ] {\cal U}\} \quad,
\label{eff4.4}
\ee
which leads to
\be
\bar\Omega^{(2m-1)} =
{2^{m-1}\over (m-1)!}
\sum_{p=0}^{m-1} (-1)^p
(m-p-1)!\, (2m-1)\cdots (2m-p) \text{Tr}
\{ \mathcal{S} \left [ (W^{p} ({\cal U}^2)^{m-p-1}) \right ] {\cal U}\} \ .
\label{eff4.5}
\ee
Now, recalling the expression of the Beta function,
\be
B(l,s) = \int_0^1 dt\, t^{l-1} (1-t)^{s-1} = {(l-1)! (s-1)! \over (l+s-1)!}
\quad,
\label{eff4.6}
\ee
we see that, renaming $p\to (m-p-1)$, $\Omega$ may be written in the form
\be
\begin{array}{rl}
\bar\Omega^{(2m-1)}
& \displaystyle
=
{2^{m-1}\over (m-1)!}
\sum_{p=0}^{m-1} (-1)^{m-p-1}
(p)! (2m-1)\cdots (m+p+1) \text{Tr}
\{ \mathcal{S} \left [ W^{m-p-1} ({\cal U}^2)^{p} \right ] {\cal U}\}
\\[0.4cm]
& \displaystyle
=
2^{m-1} {(2m-1)!\over (m-1)!\, m!}
\sum_{p=0}^{m-1} (-1)^{m-p-1}
{ p! \, m! \over (m+p) !} \text{Tr}
\{ \mathcal{S} \left [ (W^{m-p-1} ({\cal U}^2)^{p}) \right ] {\cal U}\}
\\[0.4cm]
& \displaystyle
=
(-1)^{m-1} 2^{m-1} {(2m-1)!\over (m-1)!\, m!}
\sum_{p=0}^{m-1}
\int_0^1 dt\, m\, t^{m-1} (t-1)^{p} \text{Tr}
\{ \mathcal{S} \left [ W^{m-p-1} ({\cal U}^2)^{p} \right ] {\cal U}\}
\end{array}
\label{eff4.7}
\ee
The reader will recognise that the integral in (\ref{eff4.7}) is
\emph{formally}
identical to that giving the expression of the Chern-Simons form
\cite{Che.Sim:74}
$\Omega^{(2m-1)}$ of the Chern character $ch_m$ which are relevant in the
theory
of non-abelian anomalies (see \cite{Sto:86} and
references therein).
This means that if we know the coefficients which determine the terms for the
Chern-Simons forms of various orders (see, \emph{e.g.}
\cite[\S 10.13]{Azc.Izq:95}), we also know the cocycles $\bar\Omega$ in
(\ref{eff4.4}), (\ref{eff4.5}) or (\ref{eff4.7}) and viceversa.
This explains the similarity between the two types of
(2$m$--1)-forms. In general, the $(2m)$-form $d\Omega_{CS}^{(2m-1)}$ 
gives the Chern character $ch_m$; in
contrast, the cocycle form $\bar\Omega^{(2m-1)}$ is \emph{closed},
$d\bar\Omega^{(2m-1)}=0$, since $\Pi_{(m)}$ in eq. (\ref{eff3.7}),
(\ref{eff3.5}) is zero precisely
for the polynomials which vanish on ${\cal H}$.

We may now use (\ref{eff4.7}) to recast from it the expression of the WZW
terms calculated previously.
For $m=2,3,$ eq. (\ref{eff4.7}) gives
\be
\bar\Omega^{(3)} =
- 3! [ \text{Tr} (W {\cal U} ) - {1\over 3} Tr ({\cal U}^3) ] =
[3 k_{\alpha_1 a_3} C^{\alpha_1}_{a_1 a_2} + k_{i_1 a_3} C^{i_1}_{a_1 a_2}]
\omega^{a_1}\wedge \omega^{a_2}\wedge \omega^{a_3}\quad;
\label{eff4.8}
\ee
\be
\begin{array}{rl}
\bar\Omega^{(5)} = &
\displaystyle
2^2 {5!\over 2!\, 3!}
[ \text{Tr} (W^2 {\cal U} ) - {1\over 4} \text {Tr}
(W {\cal U}^3 + {\cal U}^2 W {\cal U}) +  {1 \over 10} \text{Tr} ({\cal U}^5)]
\\[0.4cm]
& =
(10 k_{\alpha_1 \alpha_2 a_5} C^{\alpha_1}_{a_1 a_2} C^{\alpha_2}_{a_3 a_4}
+ 5 k_{\alpha_1 i_2 a_5} C^{\alpha_1}_{a_1 a_2} C^{i_2}_{a_3 a_4}
+ k_{i_1 i_2 a_5} C^{i_1}_{a_1 a_2} C^{i_2}_{a_3 a_4})
\omega^{a_1} \wedge \dots \wedge\omega^{a_5} \ ,
\displaystyle
\end{array}
\label{eff4.9}
\ee
\ie, eqs. (\ref{eff3.8}) and (\ref{eff3.9}) which have been reordered to show
the origin of their terms.
Similarly, for $m=4$ we obtain (cf. (\ref{eff3.10})),
\be
\begin{array}{rl}
\bar\Omega^{(7)} = &
\displaystyle
-2^3 {7!\over 3!\,4!} \left [ \text{Tr} (W^3 {\cal U}) - {1\over 5}
\text{Tr} ( 2 W^2 {\cal U}^3 + W {\cal U}^2 W {\cal U} )
+ {1\over 15} \text{Tr} (3 W {\cal U}^5 )
- {1\over 35} \text{Tr} ( {\cal U}^7 ) \right]
\\[0.4cm]
& =
(35 k_{\alpha_1 \alpha_2 \alpha_3 a_7}
          C^{\alpha_1}_{a_1 a_2}C^{\alpha_2}_{a_3 a_4} C^{\alpha_3}_{a_5 a_6}
+ 21 k_{\alpha_1 \alpha_2 i_3 a_7}
          C^{\alpha_1}_{a_1 a_2} C^{\alpha_2}_{a_3 a_4}C^{i_3}_{a_5 a_6}
\\[0.4cm]
& +
7 k_{\alpha_1 i_2 i_3 a_7}
          C^{\alpha_1}_{a_1 a_2} C^{i_2}_{a_3 a_4}C^{i_3}_{a_5 a_6}
+
k_{i_1 i_2 i_3 a_7} C^{i_1}_{a_1 a_2} C^{i_2}_{a_3 a_4} C^{i_3}_{a_5 a_6}
)
\omega^{a_1} \wedge \dots \wedge\omega^{a_7} \quad.
\end{array}
\label{eff4.10}
\ee

The (global) $(-)^{m-1} 2^{m-1} {(2m-1)!\over (m-1)!\, m!}$ factor in
(\ref{eff4.7}) came from the definition of 
$\Omega_{(p)}$ in (\ref{eff3.2}). By replacing this factor by
$\left ( {i\over 2\pi} \right )^m {1\over m!} (2\pi)$
we may adjust it so that the ${\cal U}^{2m-1}$ term has
the standard factor
$(-1)^{m-1}\left ( {i\over 2\pi} \right )^m {(m-1)!\over (2m-1)!} (2\pi)$
(\ie, ${1\over 12\pi},\ {-i\over 240 \pi^2},\ {-1\over 6720 \pi^3}$
for the 3, 5, 7 cocycles in (\ref{eff4.8}), (\ref{eff4.9}) and
(\ref{eff4.10})).

\vskip .5cm
{\bf Acknowledgements}

Two of the authors (J.~A.~de~A. and J.~C.~P.~B.) whish to thank the hospitality
of
DAMTP, where this work was started.
This paper has been partially supported by research grants from
the MEC, Spain (PB96-0756) and PPARC, UK.
J.~C.~P.~B. wishes to thank the Spanish MEC and the CSIC for an FPI grant.


\begin{thebibliography}{10}

\bibitem{DHo.Wei:94}
E.~D'Hoker and S.~Weinberg,
\newblock Phys. Rev. {\bf D50}, R6050--R6053 (1994).

\bibitem{DHo:95}
E.~D'Hoker,
\newblock Nucl. Phys. {\bf B451}, 725--748 (1995).

\bibitem{Azc.Mac.Mou.Bue:97}
J.A. de~Azc\'arraga, A.J. Macfarlane, A.J. Mountain, and J.~C.~P\'erez Bueno,
\newblock Nucl. Phys. {\bf B510}, 657--687 (1998) 
[{\it physics}/9706006]


\bibitem{Che.Eil:48}
C.~Chevalley and S.~Eilenberg,
\newblock Trans. Am. Math. Soc. {\bf 63}, 85--124 (1948).

\bibitem{Wei:68}
S.~Weinberg,
\newblock Phys. Rev. {\bf 166}, 1568--1577 (1968);
S.~Coleman, J.~Wess, and B.~Zumino,
\newblock Phys. Rev. {\bf 177}, 2239--2247 (1969);
C.~G. Callan, S.~Coleman, J.~Wess, and B.~Zumino,
\newblock Phys. Rev. {\bf 177}, 2247--2250 (1969).

\bibitem{Sal.Str:69}
A.~Salam and J.~Strathdee,
\newblock Phys. Rev. {\bf 184}, 1750--1759 (1969).



\bibitem{Kob.Nom:63}
S.~Kobayashi and K.~Nomizu,
\newblock {\em Foundations of differential geometry},
\newblock J. Wiley, I (1963); II (1969).

\bibitem{Wes.Zum:71}
J.~Wess and B.~Zumino,
\newblock Phys. Lett. {\bf 37B}, 95--97 (1971); 
E.~Witten,
\newblock Nucl. Phys. {\bf B223}, 422--432; 433--444 (1983).

\bibitem{wzgen}
Y.-S. Wu,
\newblock Phys. Lett. {\bf 153B}, 70--77 (1985);
I. Jack, D.R.T. Jones, N. Mohammedi and H. Osborn, 
\newblock Nucl. Phys. {\bf B332}, 359-379 (1990);
C.~M. Hull and B.~Spence,
\newblock Nucl. Phys. {\bf B353}, 379--426 (1991);
G.~Papadopoulos, Phys. Lett. {\bf B238}, 75-80 (1990);
E.~ Witten, Commun. Math. Phys. {\bf 144}, 189-212 (1992) (esp. Appendix);
J.M.~Figueroa-O'Farrill and S. Stanciu, Phys. Lett. 
{\bf B341}, 153-159 (1994).

\bibitem{Azc.Izq.Mac:90}
J.A. de~Azc\'arraga, J.M. Izquierdo, and A.J. Macfarlane,
\newblock Ann. Phys. (N.Y.) {\bf 202}, 1--21 (1990).


\bibitem{Car:36}
See \cite{Azc.Mac.Mou.Bue:97, GHV, Azc.Izq:95} for a detailed list of 
references on this point.


\bibitem{Rac:50}
G.~Racah,
\newblock Lincei-Rend. Sc. fis. mat. e nat. {\bf VIII}, 108--112 (1950);
\newblock {CERN-61-8 (reprinted in Ergeb. Exact Naturwiss.
  {\bf 37}, 28-84 (1965), Springer-Verlag)};
I.~M. Gel'fand,
\newblock Mat. Sbornik {\bf 26}, 103--112 (1950);
B.~Gruber and L. O'Raifeartaigh,
\newblock J. Math. Phys. {\bf 5}, 1796--1804 (1964);
L.~C. Biedenharn,
\newblock J. Math. Phys. {\bf 4}, 436--445 (1963);
A.~M. Perelomov and V.~S. Popov,
\newblock Math. USSR-Izvestija {\bf 2}, 1313--1335 (1968); see also
\cite{Azc.Mac.Mou.Bue:97,GHV, Azc.Izq:95} for further references.

\bibitem{Hel:78}
S.~Helgason,
\newblock {\em Differential geometry, Lie groups and Symmetric Spaces},
\newblock Academic Press, 1978.

\bibitem{Hay:76}
H.~Hayashi, I.~Ishiwata, S.~Iwao, M.~Shako and S.~Yakeshita,
\newblock Ann. Phys. (N.Y.) {\bf 101}, 394-412 (1976).

\bibitem{Che.Sim:74}
S.-S. Chern and J.~Simons,
\newblock Ann. Math. {\bf 99}, 48--69 (1974).

\bibitem{Sto:86}
R.~Stora,
\newblock {\em Jaca lectures},
\newblock in J.~Abad {\it et al.} eds., {\em New perspectives in
  quantum field theories}, pp. 309--342, World Sci., 1986;
B.~Zumino,
\newblock {\em Les Houches lectures},
\newblock in S.~B. Treiman {\it et al.} eds., {\em
  Current algebra and anomalies}, pp. 361--391, World Sci., 1985;
L.~\'Alvarez-Gaum\'e and P.~Ginsparg,
\newblock Ann. Phys. {\bf 161}, 423--490 (1985).

\bibitem{GHV}
W. Greub, S. Halperin and R. Vanstone,
{\it Connections, Curvature and Cohomology}, vol. III, Acad. Press (1976)

\bibitem{Azc.Izq:95}
J.A. de~Azc\'{a}rraga and J.~M. Izquierdo,
\newblock {\em Lie groups, Lie algebras cohomology and some applications in
  physics},
\newblock Camb. Univ. Press, 1995.

\end{thebibliography}
\end{document}